\documentclass[%
 reprint,
nofootinbib,
 amsmath,amssymb,
 aps,
]{revtex4-1}

\usepackage[utf8]{inputenc}
\usepackage{dcolumn}
\usepackage{bm}
\usepackage{physics}
\usepackage{bbold}
\usepackage{multirow}
\usepackage{makecell}
\PassOptionsToPackage{hyphens}{url}\usepackage[hidelinks]{hyperref}
\usepackage{graphics}

\newcommand{\q}[1]{`#1'}

\newcommand{\dagg}[1]{{#1}^{\dagger}}
\newcommand{\simFMS}{~\mathrel{\overset{\text{FMS}}{\scalebox{2}[1]{$\sim$}}}~}
\renewcommand{\tr}[1]{\mathrm{tr}~#1}

\newcommand{\higgsino}{\widetilde{H}}
\newcommand{\wino}{\widetilde{W}}
\newcommand{\ino}[1]{\widetilde{#1}}

\newcommand{\slepton}{\widetilde{L}}
\newcommand{\rselectron}{\widetilde{\bar{e}}}
\newcommand{\lselectron}{\widetilde{e}}
\newcommand{\rsneutrino}{\widetilde{\bar{\nu}}}
\newcommand{\lsneutrino}{\widetilde{\nu}}

\newcommand{\lepton}{L}
\newcommand{\relectron}{\bar{e}}
\newcommand{\lelectron}{e}
\newcommand{\rneutrino}{\bar{\nu}}
\newcommand{\lneutrino}{\nu}

\newcommand{\rslepton}{\ino{\bar{\lambda}}}
\newcommand{\rlepton}{\bar{\lambda}}

\newcommand{\custsym}{SU(2)_{\text{G}}}

\begin{document}

\title{The manifestly gauge-invariant spectrum of the Minimal Supersymmetric Standard Model}

\author{Axel Maas}
  \email{axel.maas@uni-graz.at}
\author{Philipp Schreiner}%
 \email{philipp.schreiner@oeaw.ac.at}
 \altaffiliation[Current affiliation: ]{Institut f\"ur Hochenergiephysik der \"Osterreichischen Akademie der Wissenschaften, 1050 Wien}
 \altaffiliation[]{{Atominstitut, Technische Universit\"at Wien, 1020 Wien}}

\affiliation{
Institute of Physics, NAWI Graz, University of Graz, Universitätsplatz 5, 8010 Graz, Austria
}

\date{\today}

\begin{abstract}
  Formal field theory requires, even in the presence of a Brout-Englert-Higgs effect, to maintain manifest non-perturbative gauge invariance. The Fr\"ohlich-Morchio-Strocchi mechanism allows nonetheless an augmented perturbative treatment. We perform such an augmented tree-level analysis for the minimal supersymmetric standard model. We find that, as for the standard model, corrections to standard perturbation theory are only sub-leading.
\end{abstract}

\maketitle

\section{Introduction}
  \label{sec:introduction}

There is an odd situation in the treatment of the Brout-Englert-Higgs effect (BEH). On the one hand, the standard textbook approach \cite{Bohm:2001yx} is tremendously successful \cite{pdg}. On the other hand, deep theorems in formal quantum field theory are at odds with the standard approach. Most notably among those is probably Elitzur's theorem \cite{Elitzur:1975im}, which states that a gauge symmetry cannot break spontaneously. In fact, from the earliest days of perturbative treatments of the BEH effect \cite{Lee:1974zg} it was recognized that it may be a manifestly gauge-dependent phenomenon \cite{Frohlich:1980gj,Frohlich:1981yi,Englert:2004yk,Englert:2014zpa}, a consequence of a very useful gauge choice \cite{Lee:1974zg,Maas:2012ct}. As such, it should not affect physical observables, which would rather need to be built from manifestly, non-perturbatively gauge-invariant operators\footnote{We note in passing that perturbative BRST is insufficient for this purpose, as it is broken \cite{Fujikawa:1978fu} by the Gribov-Singer ambiguity \cite{Gribov:1977wm,Singer:1978dk} in any non-Abelian gauge theory.} \cite{Frohlich:1980gj,Frohlich:1981yi,Banks:1979fi}. However, these are composite, and thus would be rather associated with bound states. That appears very much at odds with the highly successful phenomenology \cite{pdg} of the standard model, which is based on a perturbative treatment (PT) using elementary fields to describe observable particles \cite{Bohm:2001yx}.

This situation is relieved by the Fr\"ohlich-Morchio-Strocchi (FMS) mechanism \cite{Frohlich:1980gj,Frohlich:1981yi}. It explicitly demonstrates that the success of PT is not accidental, but rather a consequence of the structure of the standard model (SM). It shows that the correct treatment using composite operators reduces, up to small corrections, to the one of PT. The latter can be accounted for\footnote{Up to genuinely non-perturbative effects like electroweak instantons, which are supposedly small in the standard model \cite{Shifman:2012zz}.} by combining the BEH effect, the FMS mechanism and PT to construct an augmented perturbation theory (APT) \cite{Maas:2017wzi,Maas:2020kda,Dudal:2020uwb}, which can be applied directly to the manifestly gauge-invariant operators. The underlying principles and construction were tested extensively using lattice gauge theory, and have been confirmed, see \cite{Maas:2017wzi,Maas:2023emb} for a review. Thus, APT will also be the main technical approach here, and is therefore briefly introduced in section \ref{sec:apt} for a general theory. There, with the technical aspects in hand, we will also reiterate in detail on how it can appear that spontaneous electroweak symmetry breaking is experimentally established, while at the same time actually being a gauge choice.

In the SM, the remaining deviations between PT and APT are small and depend strongly on kinematical details. So far, they appear to be below experimental sensitivity \cite{Jenny:2022atm,Maas:2018ska,Maas:2020kda,Dudal:2020uwb,Maas:2023emb,Maas:2023nsa,Maas:2022gdb}. However, this is expected to change in the future, e.\ g.\ at future colliders, or perhaps already at the high-luminosity LHC \cite{Maas:2023emb}.

Beyond the SM (BSM), the differences between PT and APT may no longer be small and quantitative. Indeed, a large class of theories has been found to show \emph{qualitative} differences between PT and APT already at tree-level \cite{Maas:2016ngo,Maas:2017xzh,Maas:2023emb,Maas:2018xxu,Dobson:2022crf,Sondenheimer:2019idq}. They have been investigated using lattice calculations, and APT predictions always showed better qualitative agreement with the results than PT, see again \cite{Maas:2017wzi,Maas:2023emb} for a review. Yet some theories, like 2-Higgs doublet models (2HDM) \cite{Maas:2016qpu}, only show small quantitative differences, just like in the SM. So far, no general criterion has been found under which condition the differences are quantitative or qualitative, and it remains to check one theory (or class of theories) at a time.

We therefore investigate here one of the major candidates of BSM physics, supersymmetry (SUSY) \cite{Aitchison:2007fn,Weinberg:2000cr,Kalka:1997us}. More precisely, we perform a tree-level APT determination of the spectrum for the full minimal supersymmetric standard model (MSSM). We find that the MSSM, regarding the questions posed above, is dominated by its 2HDM-like subsector, and thus falls into the same category as the SM and the 2HDM. In particular, we find that the superpartner sector does not actively affect the FMS mechanism, nor is its spectrum qualitatively different from the one predicted by PT. Given the results in \cite{Maas:2017xzh,Sondenheimer:2019idq}, this was far from obvious.

To make the necessary steps explicit, we will build up the MSSM sector by sector. Readers, who are only interested in the full MSSM, can find the results in section \ref{subsec:generalization}. Before that, we treat a supersymmetric theory with only a Higgs and an SU(2) gauge sector in section \ref{subsec:weakHiggs} and add a lepton sector in section \ref{subsec:leptons}. Hypercharge and the strong sector are then a very straightforward extension to arrive at the complete MSSM in section \ref{subsec:generalization}. We will pay special attention to the lightest supersymmetric particle (LSP), due to its central role in the low-energy dynamics of the MSSM and dark matter \cite{Aitchison:2007fn,Weinberg:2000cr}. We will not provide a thorough introduction of the MSSM, due to its complexity, but only give a brief collection of basic formulas in section \ref{sec:mssm}. We refer to \cite{Aitchison:2007fn,Martin:1997ns} for such an introduction, whose conventions we also follow. More technical details on the calculations of this presentation can also be found in \cite{Schreiner:2022ms}.

Finally, we summarize our findings in section \ref{sec:summary}. We also add a brief appendix \ref{a:susy}, in which we give some general arguments on why SUSY does not affect the FMS mechanism.

\section{Augmented perturbation theory}\label{sec:apt}

APT is a straightforward extension of PT \cite{Maas:2017wzi,Maas:2023emb,Dudal:2020uwb,Maas:2020kda,Egger:2017tkd,Maas:2012tj,MPS:unpublished}. APT starts, in contrast to PT, by formulating the desired matrix element in the form of manifestly and non-perturbatively gauge-invariant operators. In a non-Abelian gauge theory, local operators of this type are necessarily composite. Such operators can only carry global quantum numbers. They are therefore necessarily distinct from the asymptotic states of PT, which are made of asymptotically separated particles carrying gauge indices.

Consider an uncharged scalar in the SM. The simplest operator will be built from the Higgs doublet field $\phi$, e.\ g.\ $\phi^\dagger(x)\phi(x)$. The simplest matrix element will be the connected 2-point correlation function, the propagator.

The next step in APT is, just as in PT, to choose a suitable gauge. Suitable in the case of a theory with BEH effect is a gauge with non-zero Higgs vacuum expectation value (vev) \cite{Lee:1974zg}, e.\ g.\ 't Hooft gauge. After that, the FMS mechanism is applied. This happens by rewriting the matrix element of the gauge-invariant, composite operator by explicitly splitting the Higgs field in its vev and the fluctuations, i.\ e.\ $\phi=vn+\eta$, where $v$ is the vev, $n$ a unit vector fixed by the gauge choice, and $\eta$ the fluctuation field. The FMS mechanism thus yields the following identity for the propagator of the composite operator
\begin{align}
    \left\langle (\phi^\dagger\phi)(x)(\phi^\dagger\phi)(y)\right\rangle&=v^2\left\langle(n\eta^\dagger(x))(n^\dagger\eta(y))+x\leftrightarrow y\right\rangle\nonumber\\
    &+v\left\langle(n^\dagger\eta(x))(\eta^\dagger\eta)(y)+x\leftrightarrow y\right\rangle\nonumber\\
    &+\left\langle(\eta^\dagger\eta)(x)(\eta^\dagger\eta)(y)\right\rangle.\label{apt:prop}
\end{align}
\noindent Upon choosing $n$ in the real two-direction, the first term becomes the same matrix element as appears in PT. It still has no open gauge index, as the direction $n$ carries a gauge index, contracted with the one of the fluctuation field $\eta$. Its quantum numbers are therefore still the ones of a non-gauge scalar. This feature is generic in all FMS constructions.

The other terms of  (\ref{apt:prop}) in this FMS sum are new in APT. They ensure explicit gauge invariance to all orders, which the PT term alone cannot provide beyond tree-level \cite{Maas:2020kda,Dudal:2020uwb}. Neglecting the other terms, and expanding the first term perturbatively, yields that the composite propagator is given by PT to all orders in the coupling constants and to leading order in the FMS sum. In particular, the pole position, and therefore mass and width of the composite object $\phi^\dagger\phi$, coincide to all orders in PT with those of the elementary field $h\equiv \sqrt{2}\mathrm{Re}\left(n^\dagger\eta\right)$. This has been extended \cite{Dudal:2020uwb,Maas:2020kda} by also including the remaining terms in the FMS sum, which in PT do not alter the mass or width, and yield only small changes otherwise.

To emphasize that these are small corrections, we introduce a special notation: In case a composite object behaves like the tree-level mass eigenstate of regular PT in leading order, we will write 
    \begin{equation}
    \label{eq:definition_simFMS}
        \phi^\dagger(x)\phi(x) = v^2 + \sqrt{2}vh(x)+ (\dagg\eta\eta)(x) \simFMS vh(x),
    \end{equation}  
which strips off all further contributions as well as constants other than prefactors of $v$. A constant, like $v^2$ here, will only appear in pure scalars. The neglected part will be otherwise of order $\norm{\eta}/v$, a dimensionless quantity. I.\ e.\ of order of the size of the quantum fluctuations compared to the vev. This quantity needs anyhow to be small to make an expansion about the vev sensible, which is also the starting point of PT. In lattice simulations of an isolated Higgs-$W$/$Z$ system of the SM it is indeed found that this fluctuation measure is small on average \cite{Maas:2012ct}.

In the context of this paper we stay with the tree-level results, as we are only interested in the mass spectrum. The arguments in \cite{Dudal:2020uwb,Maas:2020kda,Sondenheimer:2019idq} why the mass spectrum will not be altered by the additional terms in the FMS sum carry over verbatim from the SM to the MSSM: The additional terms in (\ref{apt:prop}) are basically 2-to-2 scattering process. A pole will only arise for an $s$-channel exchange. To have a different pole will require an $s$-channel resonance with the same quantum numbers. In the SM, and also the MSSM, no such particles exist, i.\ e.\ additional particles of the same quantum numbers with a different mass. Thus, at most the same pole as at tree-level is possible. The same argument holds true for composite operators with different quantum numbers, as has been checked on the lattice for a restricted standard model \cite{Wurtz:2013ova}. However, this is not true in general theories \cite{Dobson:2025kcx}, and the (MS)SM and 2HDM are special in this respect.

Furthermore, we will only consider propagators which have a single energy scale, the four-momentum $p$. Hence, on dimensional grounds, the expression (\ref{apt:prop}) can also be considered to be ordered by powers of $p_0^2/v^2$. In this sense, terms with less powers in $v$ can be considered as subleading, and for the purpose of the present work an ordering in powers of $v$ is feasible. This will in general no longer hold true beyond propagators \cite{MPS:unpublished}.

In the same way, PT is reproduced for all matrix elements in the SM, but not in general theories \cite{Maas:2017wzi,Maas:2023emb}. There, even the terms with highest powers in $v$ can differ qualitatively \cite{Maas:2015gma,Maas:2016ngo,Maas:2017xzh}, and thus APT and PT disagree qualitatively. This originates from group-theoretical structures, and is therefore independent of the size of any couplings. In these cases, the difference already arises for the tree-level result of the matrix element with highest power in $v$.

It now becomes evident, how it can appear as if spontaneous electroweak symmetry breaking is experimentally established, and why the standard perturbative description of electroweak physics is highly successful. The sequence here is the following:
\begin{enumerate}
    \item A gauge choice is made, like the 't Hooft gauge, which allows for a Higgs vev. This breaks the electroweak gauge symmetry explicitly, as any gauge fixing does.
    \item Dynamically, it is found that for the parameters of the standard model, the value of the Higgs vev is non-zero in this gauge. This does not need to be the case, i.\ e.\ fixing to a gauge with the possibility of a non-zero vev does not imply it to be non-zero. See e.\ g.\ \cite{Caudy:2007sf,Dobson:2022ngz} for counter examples and a detailed non-perturbative treatment.
    \item An augmented perturbative expansion around this value is made \cite{Bohm:2001yx}, yielding the original perturbative approach of Lee and Zinn-Justin \cite{Lee:1974zg}, and it is found to be in good agreement with experiment \cite{pdg,Bohm:2001yx}, indicating that non-perturbative corrections \cite{Shifman:2012zz} are small.
\end{enumerate}
Each of these steps is in itself non-trivial:
\begin{itemize}
\item[1a.] It is not necessary to fix a gauge which allows for a non-vanishing vev, but it is possible to fix to one in which it is always zero by construction \cite{Maas:2012ct}. This yields a perturbative expansion as in scalar QCD, at odds with experiment. However, non-perturbatively the results are the same, in particular gauge bosons acquire a mass, despite the vev being zero \cite{Maas:2013aia}.
\item[1b.] In special cases of theories, it is possible to eliminate the gauge degrees of freedom, and no symmetry is broken at all \cite{Maas:2017wzi,Philipsen:1996af}. Thus, a Higgs vacuum expectation value, or a criterion for electroweak symmetry breaking, cannot even be defined in lack of the Higgs field. Of course, all physical observables remain the same.
\item[1c.] Other gauge choices can have even further consequences. E.\ g.\ gauges non-linear in the Higgs field \cite{Dolan:1974gu} break the Nielsen identities \cite{Nielsen:1975fs}. Choosing such a gauge will require at the loop level more than implementing a simple pole scheme \cite{Maas:2020kda} to maintain a trivial mapping between the pole of the elementary correlator and the physical pole. 
\item[2.] That a vev is possible at all not only depends on the parameters, but actually also on the choice of gauge admitting a vev. It has been found in lattice calculations that in different gauges the formation of a vev occurs at different values of the parameters \cite{Caudy:2007sf}. I.\ e., there is a possible combination of parameters for which there is a vev in Landau gauge, while in Coulomb gauge there is none.
\item[3.] It is the success of PT, which implies that APT works:  APT yields PT plus additional contributions. The additional contributions are generically loop-suppressed due to the appearance of Bethe-Salpeter amplitudes \cite{Maas:2024hcn,MPS:unpublished}. Especially, one-scale problems show explicit suppression\footnote{However, additional threshold effects \cite{Jenny:2022atm,Maas:2018ska} still need to be understood.} in terms of scale/$v$ of the additional terms \cite{Dudal:2020uwb,Maas:2020kda,Maas:2022gdb,Jenny:2022atm,Maas:2018ska,Egger:2017tkd}. As a consequence, APT predicts that PT gives the leading contribution in the SM. This is in agreement with experimental results, which confirm that PT is a good approximation to the observations \cite{pdg}. In addition, in model theories APT provides a better description of full non-perturbative lattice data than PT, showing that if the additional contributions become leading, they are well under control \cite{Maas:2018xxu,Afferrante:2020hqe,Dobson:2025kcx}.
\end{itemize}
It is especially this last point, which is decisive, as it explains why even though it seems like spontaneous gauge symmetry breaking is an established fact, it is only a parametric suppression which make it appear so.  Thus, as an effective description, perturbation theory using 't Hooft gauge works out well. And only this chain of non-trivial aspects makes it appear as if electroweak symmetry is actually broken, providing a convenient technical starting point for calculation, without actually violating Elitzur's theorem. However, this is a consequence of the structure and parameters of the standard model, and not generally true \cite{Maas:2017wzi,Maas:2023emb}.

From a conceptual point of view, the composite operators appearing in (\ref{apt:prop}) are of the same structure as operators describing hadrons in QCD \cite{Maas:2017wzi,DeGrand:2006zz}. The only difference is that APT allows to treat them perturbatively, rather than requiring non-perturbative methods like for hadrons. In particular, this implies that aspects like the renormalizability can be treated in the same way as for hadron operators \cite{Itzykson:1980rh}. In fact, the algebraic renormalization program \cite{Piguet:1995er} has been applied already successfully to show the renormalizability of the relevant operators for the SM \cite{Dudal:2021dec}. Explicit renormalization in APT at NLO has been performed in the SM in \cite{Maas:2020kda,Dudal:2020uwb}. Thus, from the point of view of renormalizability, the operators are well-behaved.

A particular convenient consequence is that gauge-invariant composite operators are invariant under gauge transformations. This implies that so are their poles. Hence, the physical poles can never coincide with those of degrees of freedom with masses changing under a gauge transformation, like the would-be Goldstone bosons or ghosts \cite{Bohm:2001yx}. Thus, these degrees of freedom never appear in the spectrum.

On the other hand, this relegates the role of the gauge-dependent degrees of freedom, like the Higgs and the $W$/$Z$ bosons, to the same role as for quarks and gluons: They form the partons of the composite state. E.\ g., they appear in terms of PDFs as the hard degrees of freedom \cite{Maas:2022gdb,Maas:2024hcn} in interactions. But they will never appear as asymptotic states, though jets could likely be described on a similar footing as QCD jets. On the other hand, features which are entirely expressible in terms of these elementary degrees of freedom will still be expressed in this way, e.\ g.\ anomalies \cite{Maas:2017wzi}. The same is, unfortunately, also true for the hierarchy problem \cite{Maas:2017wzi,Maas:2020kda}.

\section{The MSSM}
  \label{sec:mssm}

  We will be considering the $R$-parity conserving MSSM as discussed in~\cite{Aitchison:2007fn,Martin:1997ns}. It is defined by the particle content as listed in Tab.~\ref{tab:fieldContent_MSSM}, the superpotential
\begin{equation}
    \label{eq:superpotential_MSSM}
      \begin{aligned}
        W_{\text{MSSM}} &= \ino{\bar u}\bm{y}_u\ino{Q}\cdot H_u - \ino{\bar d}\bm{y}_d\ino{Q}\cdot H_d \\
        &\quad- \rselectron\bm{y}_e\slepton\cdot H_d + \mu H_u\cdot H_d,
      \end{aligned}
\end{equation}
and the soft SUSY breaking terms
\begin{equation}
    \label{eq:softBreakingTerms_MSSM}
    \begin{aligned}
        &\mathcal{L}_{\text{MSSM}}^{\text{soft}} = -\frac{1}{2}\left[M_3\ino{g}\ino{g} + M_2\wino\wino + M_1\ino{B}\ino{B} + h.c.\right] \\
        & -\left[\ino{\bar u}\bm{a}_u\ino{Q}\cdot H_u - \ino{\bar d}\bm{a}_d\ino{Q}\cdot H_d - \rselectron\bm{a}_e\slepton\cdot H_d + h.c.\right] \\
        & - \ino{Q}^{\dagger}\bm{m}_Q^2\ino{Q} - \slepton^{\dagger}\bm{m}_L^2\slepton - \ino{\bar u}\bm{m}_{\bar u}^2\ino{\bar u}^{\dagger} -  \ino{\bar d}\bm{m}_{\bar d}^2\ino{\bar d}^{\dagger} -  \rselectron\bm{m}_{\bar e}^2\rselectron^{\dagger} \\
        & - m^2_u\dagg H_uH_u - m^2_d\dagg H_dH_d - \left[m_{ud}^2H_u\cdot H_d + h.c.\right].
    \end{aligned}
\end{equation}
All fermions are expressed as left-handed Weyl spinors, their naming follows the convention of the SM and their superpartners are denoted by a tilde above their name. The dot product is the $SU(2)$ invariant product
\begin{equation}
    \label{eq:dot_product}
    X\cdot Y \equiv X^{T}(i\sigma^2)Y = \epsilon^{ab}X_aY_b,
\end{equation}
$\bm{y}$ are the SM Yukawa matrices and $\bm{m}$ and $\bm{a}$ are $3\times 3$ matrices in flavour space as well. This information is sufficient to uniquely derive the MSSM's Lagrangian~\cite{Martin:1997ns} starting from the general form
\begin{equation}
  \label{eq:generalSUSYLagrangian}
  \begin{aligned}
      \mathcal{L} &= -\frac{1}{4}F^a_{\mu\nu}F_a^{\mu\nu} + i\ino{A}^{\dagger a}\bar{\sigma}^{\mu}(D_{\mu}\ino{A})_a\\
      &\quad+ (D_{\mu}\phi)_i^{\dagger}(D^{\mu}\phi)_i + i\ino{\phi}^{\dagger}_i\bar{\sigma}^{\mu}D_{\mu}\ino{\phi}_i\\
      &\quad- \sum_i\left|\frac{\partial W}{\partial\phi_i} \right|^2 -\frac{g^2}{2} (\dagg\phi_iT^a\phi_i)(\dagg\phi_jT_a\phi_j)\\
      &\quad- \frac{1}{2}\left[\left(\frac{\partial^2W}{\partial\phi_i\partial\phi_j}\right) \ino{\phi}_i\ino{\phi}_j + \text{h.c.}\right]\\
      &\quad- \sqrt{2}g \left[(\dagg\phi_iT^a\ino{\phi}_i)\ino{A}_a + \ino{A}^{\dagger a}(\ino{\phi}^{\dagger}_iT_a\phi_i)\right]\\
      &\quad+\mathcal{L}_{\text{soft}}.
  \end{aligned}
\end{equation}

\begin{table*}
  \caption{Field content of the MSSM. Notice that only the first family of quarks and leptons is listed explicitly. The charge assignments follow the ones used in~\cite{Aitchison:2007fn}.}
  \label{tab:fieldContent_MSSM}
  \begin{ruledtabular}
    \begin{tabular}{c c c c}
      Names & Boson & Fermion & [$SU(3)_c,SU(2)_L,U(1)_Y$] \\ \hline
      l.h. (s)quarks  & $\ino{Q}=(\ino{u},\ino{d})$ & $Q=(u,d)$ & [\textbf{3}, \textbf{2}, \hphantom{-}$\frac{1}{3}$] \\
      r.h. up (s)quark & $\ino{\bar u}$        & $\bar u$ & [$\bm{\bar{3}}$, \textbf{1}, -$\frac{4}{3}$] \\
      r.h. down (s)quark & $\ino{\bar d}$        & $\bar d$ & [$\bm{\bar{3}}$, \textbf{1}, \hphantom{-}$\frac{2}{3}$] \\ \hline
      l.h. (s)leptons  & $\slepton=(\ino{\nu},\ino{e})$ & $L=(\nu,e)$ & [\textbf{1}, \textbf{2}, -1] \\
      r.h. (s)electron & $\ino{\bar e}$        & $\bar e$ & [\textbf{1}, \textbf{1}, \hphantom{-}2] \\ \hline
      \multirow{2}{*}{Higgs(inos)} & $H_u=(H_u^+,H_u^0)$ & $\higgsino_u=(\higgsino_u^+,\higgsino_u^0)$ & [\textbf{1}, \textbf{2}, \hphantom{-}1] \\
      & $H_d=(H_d^0,H_d^-)$ & $\higgsino_d=(\higgsino_d^0,\higgsino_d^-)$ & [\textbf{1}, \textbf{2}, -1]\\ \hline
       gluons, gluinos & $g$ & $\ino{g}$ & [\textbf{8}, \textbf{1},  \hphantom{-}0]\\
       W bosons, winos & $W^{\pm},W^0$ & $\wino^{\pm},\wino^0$ & [\textbf{1}, \textbf{3},  \hphantom{-}0]\\
       B boson, bino & $B$ & $\ino{B}$ & [\textbf{1}, \textbf{1},  \hphantom{-}0]
    \end{tabular}
  \end{ruledtabular}
  \end{table*}
  
 \noindent As $SU(2)$ generators we use the scaled Pauli matrices $T^a=\sigma^a/2$.
  
  \section{A pure Weak-Higgs(ino) model}
    \label{subsec:weakHiggs}

For our considerations and the goal to establish APT in the MSSM, the dominant structures and mechanisms emerge from the weak-Higgs(ino) sector. It is therefore instructive to restrict to this subsector at first, and to add the other subsectors piece by piece until reaching the complete MSSM in section \ref{subsec:generalization}.

\subsection{Setup}

We now consider the MSSM subsector consisting only of the two Higgs $H_{u,d}$ and their superpartners $\higgsino_{u,d}$, and the $SU(2)_L$ gauge supermultiplets of the $W$ bosons together with the winos $\wino$.  Thus, the superpotential~\eqref{eq:superpotential_MSSM} contains only the last term. This yields the scalar potential part
\begin{align}
        &V(H_d,H_u) = (|\mu|^2 + m_u^2)\dagg H_uH_u + (|\mu|^2 + m_d^2)\dagg H_dH_d \nonumber \\
        &\quad+ \left(m^2_{ud} H_u\cdot H_d + h.c.\right) \label{eq:scalarPotential}\\ 
        &\quad+\frac{g^2}{8}(\dagg H_dH_d - \dagg H_uH_u)^2 + \frac{g^2}{2}(\dagg H_dH_u)(\dagg H_uH_d). \nonumber
    \end{align}
The parameter $m_{ud}^2$ can always be chosen to be real by redefining either of the Higgs fields to absorb its phase~\cite{Gunion:1984yn}. The parameters $\mu$ and $M_2$ are chosen real to avoid additional CP violations~\cite{Aitchison:2007fn,Martin:1997ns}.

The scalar potential \eqref{eq:scalarPotential}  can be re-expressed in terms of the bidoublet~\cite{Grzadkowski:2010dj} 
\begin{equation}
    \label{eq:definition_bidoublets_MSSM}
    \begin{aligned}
      H &\equiv (H_u, -H_d) =     
        \begin{pmatrix}
            H_u^+ & -H_d^0\\
            H_u^0 & -H_d^-
        \end{pmatrix},
    \end{aligned}
\end{equation}
containing both Higgs doublets. Similarly, the $\higgsino_u$ and $\higgsino_d$ are combined into a bidoublet $\higgsino$. The Higgs potential then reads
\begin{align}
    V(H) &= \tr{\dagg HHM} -2m^2_{ud}~\Re\det\dagg H - \frac{g^2}{2}\det\dagg HH \nonumber\\
    &\quad + \frac{g^2}{8}(\tr{\dagg HH})^2,\label{eq:higgs_potential}
\end{align}
with the mass matrix $M= \text{diag}(m^2_u+\mu^2,m^2_d+\mu^2)$. From this form we can read off that, so long as $m^2_u=m^2_d\equiv m^2$, the potential $V(H)$ is not only invariant under $SU(2)_L$ transformations but exhibits a second \emph{global} symmetry. This symmetry will play a central role later on, similarly to a corresponding global symmetry in the SM \cite{Maas:2017wzi}.

In total, $V(H)$ is then invariant under 
\begin{align}
    H\rightarrow H' &= L(x)H\dagg R & L(x)\in SU(2)_L,~R\in\custsym,
\end{align}
where the application from the right acts like a Higgs flavour symmetry. Note that allowing $m_u^2\neq m_d^2$ will at most break the $\custsym$ to a $U(1)$ subgroup, because the diagonal matrix $M$ can always be written in terms of $\sigma^3$ and unity.

The importance of this $U(1)$ group becomes clear after rewriting the kinetic Higgs term using the bidoublet as well:
\begin{equation}
    \label{eq:higgs_kinetic_lagrangian}
    \begin{aligned}
        \mathcal{L}_{\text{kin}}^{\text{Higgs}} &= \frac{1}{2}\tr{\left(\partial^{\mu}H - igW^{\mu}_a\frac{\sigma^a}{2}H \right)^{\dagger}\left(\partial_{\mu}H - igW_{\mu}^a\frac{\sigma^a}{2}H \right)}.
      \end{aligned}
\end{equation}
We see that a global $U(1)\subset\custsym$ transformation,
\begin{equation}
    H \rightarrow H' = H\exp (i\alpha\frac{\sigma^3}{2}),
\end{equation}
corresponds to transformations $H_{u,d}\rightarrow \exp(\pm i\alpha/2)H_{u,d}$. It is this $U(1)$ subgroup of $\custsym$ which is eventually gauged to create hypercharge, as in the SM \cite{Shifman:2012zz,Maas:2017wzi}, and persists as the $U(1)_{\text{EM}}$ group, even after introducing the $B$ field via an additional term $ig'B_{\mu} H\frac{\sigma^3}{2}$ in the kinetic term above. This will be postponed until section \ref{subsec:generalization}.

Even for $m_u^2= m_d^2$, however, in general two different vevs emerge from the Higgs potential breaking $\custsym$. To avoid this, we choose $m_{ud}^2 = \mu^2+m^2$. This choice creates flat directions and not the correct phenomenology. However, it will be lifted later and only serves as a technical auxiliary for the moment. 

The Lagrangian of the $\custsym$ preserving MSSM weak-Higgs(ino) sector is then
\begin{align}
    \mathcal{L}_{\text{WH}} &= -\frac{1}{4}W^a_{\mu\nu}W_a^{\mu\nu} + i\wino^{\dagger a}\bar{\sigma}^{\mu}(D_{\mu}\wino)_a \nonumber \\
    &+ \tr{(D_{\mu}H)^{\dagger}(D^{\mu}H)} + \tr{i\higgsino^{\dagger}\bar{\sigma}^{\mu}D_{\mu}\higgsino}\nonumber \\
    &- \frac{g}{\sqrt{2}}  \left[\tr{\dagg H\sigma^a\higgsino}\wino_a + \wino^{\dagger a}\tr{\higgsino^{\dagger}\sigma^aH}\right] \label{eq:custSymm_Lagrangian}\\
    &+ \mu\left[\det\higgsino + \text{h.c.}\right] - \frac{M_2}{2}\left[\wino^a\wino^a + h.c.\right] - V(H) \nonumber \\
    V(H) &= (\mu^2+m^2)\left[\tr{\dagg HH} -2~\Re\det\dagg H\right] \nonumber \\
    &+ \frac{g^2}{8}\tr{\dagg HH}^2 - \frac{g^2}{2}\det\dagg HH.\label{eq:custSymm_potential}
\end{align}

\subsection{Tree-level Spectrum}
    \label{subsec:EW_TreeLevel}

To judge whether there is a qualitative difference between regular PT and APT, we first calculate the tree-level spectrum using PT. For that, we construct a basis which makes later comparison to APT particularly easy.

In a suitable gauge, here 't Hooft gauge
\begin{equation}
    \label{eq:tHooftGauge}
    C^a = \partial_{\mu}W^{\mu}_a + g\xi\Im\tr{V^{\dagger}\sigma^aH},
\end{equation} 
the neutral components of the Higgs acquire the same (real) vev $v$. Employing APT, we utilize the split of $H$ in vev $V$ and fluctuation field $\eta$
\begin{equation}
    \label{eq:mssm_weakHiggs_vevExpansion}
    \begin{aligned}
        H \rightarrow V + \eta,\quad\quad V = -v(i\sigma^2).
    \end{aligned}
\end{equation}
As noted, this yields a flat direction, and thus the value of $v$ is not fixed. It is therefore treated as a free parameter until section \ref{subsec:generalization}, where the remainder of the MSSM will provide a unique value. The vev $V$ is not invariant under the full $SU(2)_L\times \custsym$ group but only under the diagonal subgroup $L_{\text{R}}V\dagg R = V$ with $L_{\text{R}} = (-i\sigma^2)R(i\sigma^2)$.

All fields will fall into multiplets of this diagonal subgroup, which will be denoted $SU(2)_m$. To make this process as transparent as possible it is useful to introduce the basis $b_i = \sigma^i(i\sigma^2)$, $(i=0,1,2,3)$, which is orthonormal with respect to the scalar product $\expval{x,y} \equiv \frac{1}{2}\tr{\dagg x y}$. Any bidoublet $Y$ can then be expressed in terms of this basis and the field bilinears $y^i$ via
\begin{equation}
    \label{eq:definition_bilinear}
    \begin{aligned}
        Y = \left(
            \begin{matrix}
                Y_2^+ & -Y_1^0\\
                Y_2^0 & -Y_1^-
            \end{matrix}
        \right) =y^ib_i\\
        y=-\frac{1}{2} \left(
            \begin{smallmatrix}
                Y_1^0+Y_2^0\\
                Y_1^-+Y_2^+\\
                i\left(-Y_1^-+Y_2^+\right)\\
                Y_1^0-Y_2^0
            \end{smallmatrix}
        \right).
    \end{aligned}
\end{equation}
The 0-component of such a vector transforms as an $SU(2)_m$ singlet, $y^0\rightarrow y'^0 = y^0$, and the remaining three as a triplet, $y^a\rightarrow y'^a = T(L_R)^{ab}y^b$. Here, $T(L_R)$ is the adjoint $SU(2)$ matrix induced by the rotation $L_R$.

In particular, we can write the Higgs fluctuation field $\eta = \zeta^ib_i$ and the Higgsino bidoublet $\higgsino=\ino{\zeta}^ib_i$ in this basis. Inserting this into the Lagrangian~\eqref{eq:custSymm_Lagrangian} yields a mass $m_W^2 \equiv g^2v^2$ for the gauge bosons as well as the mass terms
\begin{align}
    \mathcal{L} &\supset - \frac{1}{2} \left[2(\mu^2+m^2) + m^2_W\right](2\Re \zeta^a)^2 \nonumber \\
    &\quad - \frac{2(\mu^2+m^2)}{2} (2\Im \zeta^0)^2 - \xi \frac{m^2_W}{2}(2\Im \zeta^a)^2 \label{eq:custSymm_massTerms}\\
    &\quad + \sqrt{2}gv \ino{\zeta}^a\wino_a + \mu \left[\ino{\zeta}^0\ino{\zeta}^0  - \ino{h}^a\ino{\zeta}^a \right] - \frac{M_2}{2}\wino^a\wino^a + h.c.\nonumber
\end{align}
The notation $\supset$ isolates from the Lagrangian those terms, which are relevant in the following, i.\ e.\ in the calculation of tree-level (A)PT expressions, except for the trivial kinetic terms. Basically, these are all the terms necessary to determine the tree-level propagators needed.

Continuing, the spectrum contains thus a massless scalar field $h^0 \equiv 2\Re \zeta^0$, a pseudoscalar $A^0\equiv 2\Im \zeta^0$ of mass $m^2_{A^0}\equiv 2(\mu^2+m^2)$, and a mass-degenerate scalar triplet $H^a\equiv 2\Re \zeta^a$ of mass $m^2_H=m^2_A + m^2_W$. The remaining fields are the would-be Goldstone bosons $G^a\equiv 2\Im \zeta^a$ which are also an $SU(2)_m$ triplet and have the gauge-parameter dependent mass $m^2_G=\xi m^2_W$. We can further build linear combinations of the members of each triplet to get eigenstates of the $(T^3)^{ab} = i\epsilon^{3ab}$ operator. We find that $H^{\pm}\equiv (H^2\pm iH^1)/\sqrt{2}$ and $H^0\equiv H^3$ are eigenstates of definite $T^3=\pm 1,0$ and analogously for $G^a$ and $W^a$. While the notation does not make it obvious, each particle carries an open gauge index.

So far, the results are essentially analogous to the 2HDM \cite{Branco:2011iw}. For the superpartners we find a singlet\footnote{This is one of the neutralinos~\cite{Martin:1997ns}.} $\ino{N}^{0}_3 \equiv \sqrt{2}i\ino{\zeta}^0$ of mass $\mu$ and two triplets $\ino{\chi}^{a}_{1,2}$ with different masses. The latter are linear combinations of $\sqrt{2}\ino{\zeta}^a$ and $\sqrt{2}\wino^a$, i.\ e.\ they are mixtures of Higgsino and wino degrees of freedom. Again, $T^3$ eigenstates can be formed giving two more neutralinos $\ino{N}^{0}_{1,2}$ and two charginos $\ino{C}^{\pm}_{1,2}$. The results are summarized in Tab.~\ref{tab:weakHiggs_spectrum_custSymm}.
\begin{table}
    \caption{Perturbative tree-level spectrum of the weak-Higgs(ino) sector with intact symmetry $SU(2)_m$. The first column contains the usual names of the mass eigenstates, the second column denotes their field content in terms of our conveniently chosen basis (where applicable).}
    \label{tab:weakHiggs_spectrum_custSymm}
    \begin{ruledtabular}
      \begin{tabular}{c c c c}
        \thead{literature \\ names~\cite{Martin:1997ns}} & \thead{content \\ (bilinear basis)} & $SU(2)_m$ & \thead{(squared) \\ mass} \\ \hline
        $W_{\mu}^{0,\pm}$ & $W_{\mu}^a$ & $\boldsymbol{3}$ & $m_W^2$\\ 
        $h^0$ & $\Re \zeta^0$ & $\boldsymbol{1}$ & $0$ \\ 
        $A^0$ & $\Im \zeta^0$ & $\boldsymbol{1}$ & $m^2_{A^0}$ \\ 
        $H^{0,\pm}$ & $\Re \zeta^a$ & $\boldsymbol{3}$ & $m^2_{A^0}+m^2_W$\\ 
        $G^{0,\pm}$ & $\Im \zeta^a$ & $\boldsymbol{3}$ & $\xi m^2_W$\\ 
        $\ino{N}^{0}_{3}$ & $\ino{\zeta}^0$ & $\boldsymbol{1}$ & $\mu$ \\ 
        $\ino{N}^{0}_{1,2},\ino{C}^{\pm}_{1,2}$ & $\ino{\zeta}^a$, $\wino^a$ & $2\times\boldsymbol{3}$ & two different\\
      \end{tabular}
    \end{ruledtabular}
    \end{table}

\subsection{Gauge-invariant Operators}

Switching to APT allows to investigate the physical spectrum. At first, we have to construct gauge-invariant operators which is straightforward using the bidoublet formulation. These are distinguished only by their $J^{PC}$ assignment, the $\custsym$ quantum number, and $R$-parity $P_R$. They are therefore necessarily distinct by their set of quantum numbers from the elementary ones given in Tab.~\ref{tab:weakHiggs_spectrum_custSymm}. In Tab.~\ref{tab:weakHiggs_gaugeInvariantOperators} suitable operators with minimal field content up to spin $J=1$ are listed. 

\begin{table}
    \caption{Gauge-invariant bound state operators with minimal field content up to spin 1 for the $\custsym$-symmetric special case of the MSSM weak-Higgs(ino) sector. The last column contains the corresponding leading order FMS contributions.}
    \label{tab:weakHiggs_gaugeInvariantOperators}
    \begin{ruledtabular}
      \begin{tabular}{c c c c c}
        Operator & Spin & $\custsym$ & $P_R$ & $\simFMS$ \\ \hline
        $\tr{\dagg HH}$ & 0 & \textbf{1} & +1 & $v \Re\zeta^0$ \\
        $\Im\det H$ & 0 & \textbf{1} & +1  &  $v \Im\zeta^0$\\
        $\tr{\dagg HH\sigma^A}$ & 0 & \textbf{3} & +1 & $vc^{Aa}\Re \zeta^a$ \\
        $\tr{H^{\dagger}\higgsino}$ & $\frac{1}{2}$ & \textbf{1} & -1 & $v\ino{\zeta}^0$\\
        $\tr{H^{\dagger}\higgsino\sigma^A}$ & $\frac{1}{2}$ & \textbf{3} & -1 & $vc^{Aa} \ino{\zeta}^{a}$ \\
        $\tr{\dagg H\sigma^aH\sigma^A}\wino_a$ & $\frac{1}{2}$ & \textbf{3} & -1  & $v^2c^{Aa}\wino^a$ \\
        $\tr{\dagg HD_{\mu}H\sigma^A}$ & 1 & \textbf{3} & +1 & $v^2c^{Aa}W_{\mu}^a$ \\
      \end{tabular}
    \end{ruledtabular}
  \end{table}
  
  There are both scalar and fermionic $\custsym$ singlets and triplets. Applying APT yields at highest power of $v$ the operators listed in the last columns of table \ref{tab:weakHiggs_gaugeInvariantOperators}. To illustrate the process on how to arrive at the table entries, consider the operator $\tr{\dagg HH}$: As explained in Sec.~\ref{sec:apt}, every Higgs field in the operator is split into vev and fluctuation field. In the bidoublet formulation, this amounts to separating $H$ into vev $V$ and fluctuation field $\eta$ according to Eq.~\eqref{eq:mssm_weakHiggs_vevExpansion}. Furthermore, we express $\eta$ in the basis of Eq.~\eqref{eq:definition_bilinear}. Tree-level APT then yields
  \begin{align*}
     &\tr{\dagg HH} = \tr\left[\dagg{\left(V+\zeta^ib_i\right)}\left(V+\zeta^jb_j\right)\right]\\
     &\quad= \tr\left[\dagg{\left(-v(i\sigma_2)+\zeta^i\sigma_i(i\sigma_2)\right)}\left(-v(i\sigma_2)+\zeta^j\sigma_j(i\sigma_2)\right)\right]\\
     &\quad= \tr\left[v^2\sigma_0 - 2v\left(\Re\zeta^i\right)\sigma_2\sigma_i\sigma_2 + \zeta^{i*}\zeta^j\sigma_2\sigma_i\sigma_j\sigma_2\right]\\
     &\quad= 2v^2 - 4v\delta_{0i}\Re\zeta^i + 2\delta_{ij}\zeta^{i*}\zeta^j\\
     &\quad= 2v^2 - 4v\Re\zeta^0 + 2\zeta^{i*}\zeta^i\\
     &\quad\simFMS v\Re \zeta^0 \sim vh^0
  \end{align*}
i.e. after discarding the constant and terms that are $\norm{\zeta}/v$ suppressed, the operator reduces to the elementary scalar singlet. Likewise, the fermionic singlet operator reduces to $\ino{\zeta}^0$. 

The matrix $c^{Aa}$ in Tab.~\ref{tab:weakHiggs_gaugeInvariantOperators} has to be highlighted because it maps gauge indices $a$ to $\custsym$ (physical) indices $A$. Because of its special form
\begin{equation}
    \label{eq:cAa}
    c^{Aa} = \text{diag}(1,-1,1)
\end{equation}
this mapping is one-to-one but not trivial like in the SM~\cite{Maas:2017wzi}. This shows that the nonphysical $SU(2)_m$ triplets found in the previous section using PT map to $\custsym$ triplets and so does their mass degeneracy.

Notice that the mapping to $SU(2)_m$ eigenstates is sufficient as at tree-level the FMS mechanism is transparent to linear combinations. This allows to construct mass eigenstates, in case they differ. For example the manifestly gauge-invariant charged Higgs, which reduces to the elementary charged Higgs $H^{\pm}$ through APT, can be constructed as\footnote{Notice that the $\mp$ between the Pauli matrices is due to the special structure of~\eqref{eq:cAa} and opposite to our definition of $SU(2)_m$ $T^3$ eigenstates in Sec.~\ref{subsec:EW_TreeLevel}, like $H^{\pm}= (H^2\pm iH^1)/\sqrt{2}$ for example.}
\begin{align*}
    \tr{\dagg HH(\sigma^2\mp i\sigma^1)} \simFMS -v(4\Re\zeta^2 \pm 4i\Re\zeta^1) \sim vH^{\pm}.
\end{align*}
It is important to realize that there is no gauge-invariant operator mapping to the (unphysical) would-be Goldstone bosons $G^{0,\pm}\sim \Im \zeta^a$. They are \q{projected} out of the physical spectrum automatically. This replaces their usual removal from the physical spectrum by a BRST construction\footnote{The would-be Goldstone bosons carry a gauge-parameter-dependent mass \cite{Bohm:2001yx}. As such, the corresponding poles cannot appear as poles of a gauge-invariant composite operator, which is invariant under gauge parameter changes. This can be checked explicitly. Operators with the same global quantum numbers as the Goldstones can be constructed, as they are a scalar triplet of $\custsym$. In the SM, this quantum number channel has been investigated in lattice simulations \cite{Wurtz:2013ova}, and no single-particle states have been seen, confirming this argument. In the APT treatment here and in the 2HDM model in \cite{Maas:2016ngo} such states would appear as poles in the investigated triplet channel, but with gauge-parameter dependent masses, which is not seen. This can be traced back to the fact that APT yields order-by-order gauge(-parameter) independent results \cite{Maas:2020kda}.}.

In total, all the new Higgs, charginos and neutralinos predicted by PT have a gauge-invariant counterpart. In particular, this holds for the lightest of the neutralinos, the LSP.

  \section{Leptons}
    \label{subsec:leptons}

    \subsection{Setup}

    The next step is to include one lepton generation, i.e. a left-handed lepton $L$ with its superpartner $\slepton$ and the right-handed (s)electron $\relectron$ ($\rselectron$). For simplicity a right-handed neutrino and its superpartner is added,  $\rneutrino$ and $\rsneutrino$, respectively. The superpotential is then 
\begin{equation}
    \label{eq:Leptons_superpotential}
    W = \mu H_u\cdot H_d -y_e\rselectron\slepton\cdot H_d +y_{\nu}\rsneutrino\slepton\cdot H_u,
\end{equation}
and the soft breaking Lagrangian includes
\begin{equation}
    \label{eq:Leptons_softTerms2}
    \begin{aligned}
        \mathcal{L}_{\text{soft}} &\supset a_e\rselectron\slepton\cdot H_d - a_{\nu}\rsneutrino\slepton\cdot H_u \\
        &\quad- m^2_L\dagg\slepton\slepton - m^2_{\relectron}\dagg\rselectron\rselectron - m^2_{\rneutrino}\dagg\rsneutrino\rsneutrino,
    \end{aligned}
\end{equation}
in addition to the Higgs and Wino contributions from Sec.~\ref{subsec:weakHiggs}.

For technical simplicity, we assume for the moment
\begin{equation}
    \label{eq:parameter_degeneracy_Leptons}
    \begin{aligned}
        y_e=y_{\nu} &\equiv y\\
        a_e=a_{\nu} &\equiv a \\
        m_{\relectron}=m_{\rneutrino} &\equiv m_{\bar{\lambda}},
    \end{aligned}
\end{equation}
such that the $\custsym$ of the weak-Higgs(ino) sector does not get entirely broken yet. 

The $\relectron$ and $\rneutrino$ (as well as their superpartners) can then be put into $SU(2)_F$ flavour doublets $\rlepton \equiv (\rneutrino,\relectron)^T$ and $\rslepton \equiv (\rsneutrino,\rselectron)^T$, which yields for the Lagrangian
\begin{equation}
    \label{eq:Leptons_lagrangian_symmetric}
    \begin{aligned}
        \mathcal{L} &=  \mathcal{L}_{\text{WH}} + (D_{\mu}\slepton)^{\dagger}(D^{\mu}\slepton) + i\dagg \lepton\bar{\sigma}^{\mu}D_{\mu}\lepton \\
        &\quad+ (\partial_{\mu}\rslepton)^{\dagger}(\partial^{\mu}\rslepton) + i\rlepton^{\dagger}\bar{\sigma}^{\mu}\partial_{\mu}\rlepton\\
        &\quad-\frac{g^2}{8}\left[(\dagg\slepton\sigma^a\slepton)^2 + (\dagg\slepton\sigma^a\slepton)~\tr{\dagg H\sigma^aH} \right]  \\
        &\quad-\frac{g}{\sqrt{2}}\left[(\dagg\slepton\sigma^a\lepton)\wino_a + h.c.\right]\\
        &\quad-|y|^2(\dagg\rslepton\rslepton)(\dagg\slepton\slepton)- \left[\mu^* y\rslepton\cdot\dagg H\slepton + h.c.\right] \\
        &\quad-|y|^2 \left(H\rslepton\right)^{\dagger}\left(H\rslepton\right) - |y|^2(\slepton\cdot H)(\slepton\cdot H)^{\dagger}\\
        &\quad- \left[y(\lepton\cdot\higgsino)\rslepton + y(\slepton\cdot\higgsino)\rlepton + y(\lepton\cdot H)\rlepton + h.c.\right] \\
        &\quad-\left[a(\slepton\cdot H)\rslepton + h.c.\right] -m_L^2\dagg\slepton\slepton - m_{\rlepton}^2\dagg\rslepton\rslepton.
    \end{aligned}
\end{equation}
When (degenerate) Yukawa couplings $y$ and/or the soft breaking parameter $a$ is non-zero, the $\custsym$ and the flavour symmetry break down to a diagonal subgroup, $\custsym\times SU(2)_F \rightarrow SU(2)_{f}$.

\subsection{Tree-level Spectrum}

After the Higgs acquires its vev, $H = v(-i\sigma^2) + \eta$, the relevant parts of the Lagrangian are
\begin{align}
    \mathcal{L} &\supset - \begin{pmatrix}\slepton^T & \rslepton^{\dagger}\end{pmatrix}
        \begin{pmatrix}
            v^2y^2+m^2_L & v(a-y\mu) \\ v(a-y\mu) & v^2y^2+m^2_{\bar{\lambda}}
        \end{pmatrix}
        \begin{pmatrix} \slepton^{\dagger T} \\ \rslepton \end{pmatrix}
        \nonumber\\
        &\quad- vy \left[(\lneutrino\rneutrino + \lelectron\relectron) + h.c.\right]. \label{eq:leptons_massLagrangian}
\end{align}
The last bracket contains Dirac mass terms of the electron and neutrino. We therefore combine their left-handed and right-handed components into Dirac spinors $\psi^e \equiv (\lelectron,\dagg\relectron)^T$ and $\psi^{\nu} \equiv (\lneutrino,\dagg\rneutrino)^T$, and, since they have the same mass $vy$, we collect them into a doublet $\psi \equiv (\psi^{\nu},\psi^e)^T$. A straightforward calculation yields the mass terms
\begin{equation}
    \begin{aligned}
        \mathcal{L} &\supset -m^2_{\phi_1}\dagg\phi_1\phi_1 -m^2_{\phi_2}\dagg\phi_2\phi_2 - m_{\psi}\bar{\psi}\psi\\
        m^2_{\phi_{1,2}} &= \frac{1}{2}\left(m^2_L + m^2_{\bar{\lambda}} + 2v^2y^2 \right.\\
        &\quad\left.\pm \sqrt{ \left(m^2_L -m^2_{\bar{\lambda}}\right)^2 + 4v^2(a-y\mu)^2}\right)\\
        m_{\psi} &= vy.
    \end{aligned}
\end{equation}
The fields $\phi_{1,2}$ are linear combinations of $\slepton^{\dagger T}$ and $\rslepton$, i.e. they still contain selectrons and sneutrinos which form mass-degenerate doublets. From that we can read off that our mass spectrum consists of two scalar doublets of mass $m^2_{\phi_{1,2}}$ and a doublet of Dirac fermions of mass $m_{\psi}=vy$. However, they are \q{doublets} with respect to different symmetries, but they actually mix $SU(2)_L$ and $SU(2)_F$ rotations. E.g. $\slepton$ transforms under $SU(2)_L$ while $\rslepton$ transforms under $SU(2)_F$ and $\phi_{1,2}$ transform under neither of them. 

\subsection{Gauge-invariant Operators}

The right-handed leptons are already gauge-invariant with respect to the weak interactions. The left-handed lepton doublets are not. A gauge-invariant operator can be constructed as a composite operator of a left-handed lepton and the Higgs bidoublet. It again carries no open gauge index, and therefore is distinct from the elementary leptons.

\begin{table}
    \caption[Gauge-invariant operators in the lepton toy model]{Gauge-invariant operators in the lepton toy model and their quantum numbers.}
    \label{tab:leptonToyModel_gaugeInvariantOperators}
    \begin{ruledtabular}
      \begin{tabular}{c c c c c c}
         Operator & Spin & $\custsym$ & $SU(2)_F$ & $SU(2)_{f}$ & $P_R$ \\ \hline
         $\dagg H\lepton$ & $\frac{1}{2}$ & \textbf{2} & \textbf{1} & \textbf{2} &$+1$ \\
         $\dagg H\slepton$ & 0 & \textbf{2} & \textbf{1} & \textbf{2} &$-1$ \\
         $\rlepton$ & $\frac{1}{2}$ & \textbf{1} & \textbf{2} & \textbf{2} &$+1$ \\
         $\rslepton$ & 0 & \textbf{1} & \textbf{2} & \textbf{2} &$-1$ \\
      \end{tabular}
    \end{ruledtabular}
  \end{table}
  
  In Tab.~\ref{tab:leptonToyModel_gaugeInvariantOperators} we state the possible operators with minimal field content and both the $\custsym$ and $SU(2)_F$ multiplet structure to emphasize the point that both are important and their interplay is crucial. Nonetheless, the only remaining (global) symmetry is the diagonal symmetry $SU(2)_{f}$ and all states listed in the table are doublets with respect to that group. Both $\dagg HL$ and $\rlepton$ are Weyl spinors and can readily be combined into Dirac spinors
\begin{equation}
    \label{eq:electron_neutrino_operators_equal_vev}
    \begin{aligned}
        \Psi^e \equiv \begin{pmatrix}
            (\dagg HL)_1 \\ v(\rlepton^c)_1
        \end{pmatrix},\quad \Psi^{\nu} \equiv \begin{pmatrix}
            (\dagg HL)_2 \\ v(\rlepton^c)_2
        \end{pmatrix}.
    \end{aligned}
\end{equation}
Notice that the charge conjugation acts on the doublet of Weyl spinors as $\rlepton^c = i\sigma^2(\dagg\rneutrino,\dagg\relectron)^T = (\dagg\relectron,-\dagg\rneutrino)^T$ and that $\rlepton^c$ transforms identical to $\rlepton$ under $SU(2)_F$, resp. $SU(2)_{f}$. 

We can now use those elements to build a gauge-invariant lepton doublet\footnote{Notice that $(\dagg HL)_{1,2} = \dagg H_{u,d}L \simFMS e,\nu$, i.\ e.\ that depending on which Higgs the left-handed elementary lepton is dressed with, it either results in a physical electron or a physical neutrino.}
\begin{equation}
    \begin{aligned}
    \Psi = \begin{pmatrix}
        \Psi^{\nu} \\ \Psi^e
    \end{pmatrix}=\begin{pmatrix}
        [(\dagg HL)_2,-v\dagg\rneutrino]^T \\ [(\dagg HL)_1,~v\dagg\relectron]^T
    \end{pmatrix}\\
    \quad\simFMS v\begin{pmatrix}
        -[\lneutrino,~\dagg\rneutrino]^T \\ \hphantom{-}[\lelectron,~\dagg\relectron]^T
    \end{pmatrix} = v\begin{pmatrix}
        -\psi^{\nu} \\ \psi^e
    \end{pmatrix}
\end{aligned}
\end{equation}
which transforms like a doublet under $SU(2)_f$ and reduces to the elementary lepton \q{doublet} in tree-level APT.

The scalar partners of the leptons will also form a doublet. They are readily constructed from the operators in the table as
\begin{equation}
    \label{eq:slepton_doublet_GIOperator}
    \begin{aligned}
        \Phi_{1,2} &\equiv \alpha_{1,2}i\sigma^2(\dagg H\slepton)^* + \beta_{1,2}v\rslepton \\
        &\quad \simFMS v\left(\alpha_{1,2}\slepton^{\dagger T} + \beta_{1,2}\rslepton\right) = v\phi_{1,2}
    \end{aligned}
\end{equation}
where $\alpha$ and $\beta$ encode the relative phases obtained by diagonalizing the mass matrix in Eq.~\eqref{eq:leptons_massLagrangian}. $\Phi_{1,2}$ now truly transform as doublets, but under $SU(2)_{f}$. Hence, we once again observe that the perturbative \q{doublet} is obtained from a map of composite physical operators. In total, we have shown that the physical spectrum indeed contains a doublet of Dirac fermions $\Psi$ as well as two scalar doublets $\Phi_{1,2}$, the masses of which are the same as in PT. In the totally symmetric case the members of the doublets are again mass degenerate.

Next, we will see how those degeneracies are lifted once we turn to the more realistic case of $v_d\neq v_u$, $y_e\neq y_{\nu}$, $a_e\neq a_{\nu}$ and $m_{\relectron}\neq m_{\rneutrino}$. All those changes break both the $\custsym$ as well as the flavour symmetry from before. Consequently, $SU(2)_{f}$ will not be an exact symmetry of the theory anymore and the degeneracies of the bound states will be lifted. Via APT, this in turn explains how the elementary \q{doublets} obtain different masses in a completely gauge-invariant fashion.
    
  \section{Generalization to the whole MSSM}
    \label{subsec:generalization}

  Up until now, we intentionally kept $\custsym$ intact and many parameters degenerate in order to make the calculations of the tree-level spectra easier and to see the structure behind how the different gauge-invariant operator multiplets map to the gauge-dependent multiplets in the elementary field description. Lifting these restrictions is straightforward in the sense that the calculations done so far generalize. In Sec.~\ref{subsubsec:brokenCustodialSymmetry} we will see how explicitly breaking $\custsym$ symmetry and having non-degenerate Yukawa couplings and soft breaking terms mixes both the elementary spectra as well as the corresponding bound state operators. Including hypercharge into the description is analogous to the SM (c.f.~\cite{Maas:2017wzi}) and discussed in Sec.~\ref{subsubsec:QED}. Finally, Sec.~\ref{subsubsec:QCD} discusses the extension to quarks and hadrons as well as multiple fermion generations.

\subsection{Broken Global Symmetry}
    \label{subsubsec:brokenCustodialSymmetry}
    
So far, the two Higgs fields acquired the same vacuum expectation value. This will change now. Consider first the weak-Higgs(ino) sector. Introducing separate soft breaking masses for the two Higgs, making their off-diagonal mass an independent parameter, and letting them acquire different vevs explicitly breaks the $\custsym$ as can be seen from Eq.~\eqref{eq:higgs_potential}. We therefore expect the previously found mass eigenstates to mix with each other. A detailed discussion of this potential including the calculation of the scalar masses can, e.g., be found in~\cite*{Aitchison:2007fn}.

However, we want to take a slightly different route here. Minimizing the potential leads to two independent vevs, $v_u$ and $v_d$. We define $v^{\pm}\equiv v_u\pm v_d$ and align their directions in the bidoublet language as
\begin{equation}
  \label{eq:MSSM_higgsVEV_unequal}
  H = \begin{pmatrix}
    0 & -v_d \\ v_u & 0
\end{pmatrix} + \eta = \frac{v^+}{2}(-i\sigma^2) + \frac{v^-}{2}\sigma^1 + \eta.
\end{equation}
Without loss of generality, we choose $v_u>v_d$ and we define $\tan\beta \equiv v_u/v_d$.

After inserting the split~\eqref{eq:MSSM_higgsVEV_unequal} into the Lagrangian we find the gauge-boson mass $m_W^2 = \frac{g^2}{2}(v_u^2+v_d^2)$. After expressing the fluctuation fields again via the bilinears defined in Eq.~\eqref{eq:definition_bilinear}, i.e. $\eta=\zeta^ib_i$, and replacing them by their literature names afterwards (c.f. Tab.~\ref{tab:weakHiggs_spectrum_custSymm}), we can write the scalar mass matrix in block-diagonal form
\begin{align}
    \mathcal{L} &\supset \frac{1}{2}\begin{pmatrix}
        h^0 & H^0
    \end{pmatrix}
    \Lambda_1
    \begin{pmatrix}
        h^0 \\ H^0
    \end{pmatrix} +
    \frac{1}{2}\begin{pmatrix}
        A^0 & G^0
    \end{pmatrix}
    \Lambda_2
    \begin{pmatrix}
        A^0 \\ G^0
    \end{pmatrix} \nonumber \\
    &\quad+
    \begin{pmatrix}
        H^{+} & G^{+}
    \end{pmatrix}
    \Lambda_3
    \begin{pmatrix}
        H^{-}\\ G^{-}
    \end{pmatrix}. \label{eq:massLagrangian_weakHiggs_nonSymmetric}
  \end{align}
  We see that the fields only mix pair-wise as compared to the previous (symmetric) case.
  
  Solving the eigenvalue problem yields the explicit mixing for the pseudoscalar and charged scalars
  \begin{align*}
      A^{0'} &= \frac{1}{\sqrt{2}}\left[(\cos\beta+\sin\beta)A^0 + (\sin\beta-\cos\beta)G^0\right]   \\
      H^{-'} &= \frac{1}{\sqrt{2}}\left[i(\cos\beta+\sin\beta)H^{-} + (\sin\beta-\cos\beta)G^{-}\right].
  \end{align*}
  Primed fields correspond to mass eigenstates of the non-$\custsym$-symmetric case. For $v_d=v_u$, or $\cos\beta=\sin\beta$, the relations reduce to the previous pseudoscalar $A^0$ and charged scalars $H^-$, even though the entire model does not approach the fully symmetric case in that limit.  This becomes apparent when looking at the neutral Higgs scalars $h'$ and $H^{0'}$ where the mixing is not just because $v_u$ and $v_d$ are different but also due to the newly introduced parameters, and  $m_u^2$, $~m_d^2$, $~m_{ud}^2$ are all independent. Nevertheless, we know from~\eqref{eq:massLagrangian_weakHiggs_nonSymmetric}, that they will be linear combinations of the previously found fields $h^0$ and $H^0$.
  
  For the gauge-invariant operators we find that the ones corresponding to $A^0$ and $H^{\pm}$ from before (c.f.~Tab.~\ref{tab:weakHiggs_gaugeInvariantOperators}) now automatically reduce to the new mass eigenstates, i.e.
  \begin{align*}
      \Im\det H &\simFMS v^+A^0 + v^-G^0 \sim A^{0'}\\
      \tr{\dagg HH(\sigma^2\mp i\sigma^1)} &\simFMS v^+H^{\pm} \pm i v^-G^{\pm} \sim H^{\pm'}.
  \end{align*}
  As already mentioned, this is not the case for $h^0$ and $H^0$. However, we can express $h^0$ and $H^0$ via the operators
  \begin{align*}
      v^+\tr{\dagg HH} - v^-\tr{\dagg HH\sigma^3} &\simFMS v_dv_uh^0\\
      v^-\tr{\dagg HH} - v^+\tr{\dagg HH\sigma^3} &\simFMS v_dv_uH^0
  \end{align*}
  which in leading order correspond to the elementary fields of the fully symmetric case. Appropriate linear combinations then reduce to the corresponding primed fields.

  Next, we turn to the Higgsinos and winos: The relevant parts of the Lagrangian are unchanged, however, the new vev introduces additional mixing terms
  \begin{align*}
    \mathcal{L} &\supset \frac{g}{\sqrt{2}}v^+\left[ \ino{\zeta}^a\wino_a + \wino^{a\dagger}\ino{\zeta}^{a\dagger} \right] \\
    &\quad-\frac{g}{\sqrt{2}}v^-\left[\ino{\zeta}^0\wino^3 + i\ino{\zeta}^2\wino^1 - i\ino{\zeta}^1\wino^2 + h.c.\right]  \\
      &\quad+ \mu \left[\ino{\zeta}^0\ino{\zeta}^0 + \ino{\zeta}^{0\dagger}\ino{\zeta}^{0\dagger} - \ino{\zeta}^a\ino{h}^a - \ino{\zeta}^{a\dagger}\ino{\zeta}^{a\dagger} \right].
  \end{align*}
  This can again be brought into block diagonal form in the basis of Tab.~\ref{tab:weakHiggs_spectrum_custSymm}
  \begin{align*}
    \mathcal{L} &\supset\begin{pmatrix}
        \ino{N}^0_1 & \ino{N}^0_2 & \ino{N}^0_3
    \end{pmatrix} 
    \begin{pmatrix}
        \cdot & * & *\\
        * & \cdot & * \\
        * & * & \cdot
    \end{pmatrix}
    \begin{pmatrix}
        \ino{N}^0_1 \\ \ino{N}^0_2 \\ \ino{N}^0_3
    \end{pmatrix}\\
    &\quad+ 
    \begin{pmatrix}
        \ino{C}^-_1 & \ino{C}^-_2
    \end{pmatrix}
    \begin{pmatrix}
        \cdot & * \\ * & \cdot
    \end{pmatrix}\
    \begin{pmatrix}
        \ino{C}^+_1 \\ \ino{C}^+_2
    \end{pmatrix} + h.c.
  \end{align*}
  where the off-diagonal elements $*$ are terms of the form $(v_u^2+v_d^2)^{1/2}-(v_u+v_d)/\sqrt{2}$ or $(v_d-v_u)$ and therefore vanish when $v_u=v_d$. We have thus demonstrated that the neutral (charged) fermions mix once $\custsym$ symmetry is violated. The new mass eigenstates approach the previous ones when the symmetry is restored. Once again, appropriate linear combinations of the operators found in Sec.~\ref{subsec:weakHiggs} can be used to build gauge-invariant bound states. Those in turn reduce to the elementary mass eigenstates in tree-level APT. In particular, an operator which augments the lightest of the uncharged fermions can be constructed, i.e.\ the LSP remains part of the physical spectrum.
  
  At last, we investigate the case of $v_d\neq v_u$ in the (s)lepton sector.  Additionally, we set now $y_e\neq y_{\nu}$, $a_e\neq a_{\nu}$ and $m^2_{\relectron}\neq m^2_{\rneutrino}$ but still assume that they are real.  Inserting the new split~\eqref{eq:MSSM_higgsVEV_unequal} into the lepton Lagrangian yields
  \begin{align*}
      \mathcal{L} &\supset  -\dagg\xi_1X_1\xi_1 - \dagg\xi_2X_2\xi_2 - v_dy_e\bar\psi^e\psi^e - v_uy_{\nu}\bar\psi^{\nu}\psi^{\nu}.
  \end{align*}
    We immediately see that the lepton doublet splits, with masses proportional to the different vevs. The slepton masses are currently written in the basis $\xi_1=(\lsneutrino,\dagg\rsneutrino)^T$, $\xi_2 =(\lselectron,\dagg\rselectron)^T$ with the matrices
  \begin{align*}
      X_1 &= \begin{pmatrix}
                  \left(y_{\nu}^2-\frac{g^2}{8}\right)v_u^2 + \frac{g^2}{8}v_d^2 +m^2_L & v_ua_{\nu}-\mu v_dy_{\nu} \\
                  v_ua_{\nu}-\mu v_dy_{\nu} & v_u^2y_{\nu}^2 + m^2_{\rneutrino}
              \end{pmatrix}\\
    X_2 &= \begin{pmatrix}
              \left(y_{e}^2-\frac{g^2}{8}\right)v_d^2 + \frac{g^2}{8}v_u^2 +m^2_L & v_da_{e}-\mu v_uy_{e} \\
              v_da_{e}-\mu v_uy_{e} & v_d^2y_{e}^2 + m^2_{\relectron}
          \end{pmatrix}
  \end{align*}
  which are yet to be diagonalized. It is straightforward to do so but adds nothing new apart from four different slepton mass eigenstates. We notice, however, that for the case of equal vevs and degenerate $y,a,m_{\rlepton}$, both mass matrices reduce to the ones found in the fully symmetric case which restores the mass-degenerate doublets.
  
  For the leptons, we can immediately write down composite operators
\begin{equation}
    \label{eq:electron_neutrino_operators_unequal_vev}
      \begin{aligned}
          \Psi^{e} &= \begin{pmatrix}
            (\dagg HL)_1 \\ v_u(\rlepton^c)_1
          \end{pmatrix} \simFMS v_u\psi^e\\
          \Psi^{\nu} &=\begin{pmatrix}
            (\dagg HL)_2 \\ v_d(\rlepton^c)_2
          \end{pmatrix} \simFMS v_d\psi^{\nu}
      \end{aligned}
\end{equation}
  which are essentially the lepton operators found in the fully symmetric case,~Eq.~\eqref{eq:electron_neutrino_operators_equal_vev}. Now, they merely expand with the different vevs. The slepton mass eigenstates will be linear combinations of $\lsneutrino$ and $\dagg\rsneutrino$ ($\lselectron$ and $\dagg\rselectron$, respectively) which is why it is sufficient to know that
  \begin{align*}
      (\dagg H\slepton)_1  &\simFMS v_u\lselectron & (\dagg H\slepton)_2  &\simFMS v_d\lsneutrino \\
      v_u(\dagg\rslepton)_2 &\simFMS v_u\dagg\rselectron & v_d(\dagg\rslepton)_1 &\simFMS v_d\dagg\rsneutrino.
  \end{align*}
  Those operators are all gauge-invariant and can be combined such that they match whatever form the explicit mass eigenstates have\footnote{Note that the linear combinations are formed between $(\dagg H\slepton)_1$ and $(\dagg\rslepton)_2$, i.e. with the components reversed. This is not a mistake and also present in the fully symmetric case, Eq.~\eqref{eq:slepton_doublet_GIOperator}, where this \q{mixing} is slightly hidden by $i\sigma^2$. Furthermore, the sneutrino expands with $v_d$ whereas the selectron expands with $v_u$. This is indeed opposite to their masses, which are proportional to $v_u$ and $v_d$, respectively.}.

    We conclude, that a mapping between perturbative mass eigenstates and physical composite states is possible, even for different Higgs vevs and non-degenerate couplings. The mixing of the perturbative mass eigenstates is completely parallel to the mixing of the physical composite state operators.

\subsection{Electric Charge and QED}
  \label{subsubsec:QED}
  
  As already mentioned, $U(1)_Y$ is a subgroup of $\custsym$. As a result, some fields that are assigned a hypercharge of zero in the elementary field description (e.g. $W_{\mu}^a$) acquire a non-zero electric charge also in the composite operator language. We should therefore check whether the operators we constructed carry the same electric charge as their elementary counterparts. For that it is sufficient to investigate what effect the (global) hypercharge transformations (c.f.~Tab.~\ref{tab:fieldContent_MSSM})
\begin{align*}
    H &\rightarrow H'=H~\exp(i\alpha\frac{\sigma^3}{2})\\
    L &\rightarrow L'=e^{-i\alpha/2}L,\quad \relectron \rightarrow \relectron' =e^{2i\alpha/2}\relectron
\end{align*}
of the elementary fields have on the composite operators.

The scalar and pseudo-scalar singlet operators $\tr{\dagg HH}$ and $\Im\det H$ are invariant under such transformation because of the properties of trace and determinant. Hence, they are charge neutral just like their elementary counterparts $h^0$ and $A^0$. Likewise, the LSP operator $\tr{\dagg H\higgsino}$ is charge neutral. The Higgs triplet transforms as 
\begin{align*}
    \begin{pmatrix}
        \mathcal{O}_{H^+}\\ \mathcal{O}_{H^-} \\ \mathcal{O}_{H^0}
    \end{pmatrix} &=
    \begin{pmatrix}
        \tr{\dagg HH(\sigma^2-i\sigma^1)} \\ \tr{\dagg HH(\sigma^2+i\sigma^1)} \\ \tr{\dagg HH\sigma^3}
    \end{pmatrix} \\
    &\rightarrow
    \begin{pmatrix}
        \mathcal{O}'_{H^+}\\ \mathcal{O}'_{H^-} \\ \mathcal{O}'_{H^0}
    \end{pmatrix}=
    \begin{pmatrix}
        e^{i\alpha} & & \\ & e^{-i\alpha} & \\ & & 1
    \end{pmatrix}\begin{pmatrix}
        \mathcal{O}_{H^+}\\ \mathcal{O}_{H^-} \\ \mathcal{O}_{H^0}
    \end{pmatrix},
\end{align*}
which confirms that they indeed carry electric charges of $0,\pm 1$, just like the corresponding $H^{0,\pm}$. The same can be done for the remaining triplet operators of the pure weak-Higgs sector as all of them boil down to the same rotation.

The left-handed (s)leptonic operators transform as 
\begin{align*}
    \begin{pmatrix}
        \mathcal{O}_e \\ \mathcal{O}_{\nu}
    \end{pmatrix} = 
    \dagg H\lepton &\rightarrow \begin{pmatrix}
        \mathcal{O}'_e \\ \mathcal{O}'_{\nu}
    \end{pmatrix} = 
    \begin{pmatrix}
        e^{-i\alpha} & \\ & 1
    \end{pmatrix}\begin{pmatrix}
        \mathcal{O}_e \\ \mathcal{O}_{\nu}
    \end{pmatrix}\\
    \begin{pmatrix}
        \mathcal{O}_{\widetilde{e}} \\ \mathcal{O}_{\widetilde{\nu}}
    \end{pmatrix} = \dagg H\slepton &\rightarrow
    \begin{pmatrix}
        \mathcal{O}'_{\widetilde{e}} \\ \mathcal{O}'_{\widetilde{\nu}}
    \end{pmatrix} = 
    \begin{pmatrix}
        e^{-i\alpha} & \\ & 1
    \end{pmatrix}
    \begin{pmatrix}
        \mathcal{O}_{\widetilde{e}} \\ \mathcal{O}_{\widetilde{\nu}}
    \end{pmatrix}
\end{align*}
 and therefore their charge assignment is correct as well. Notice that right-handed fields automatically adopt their hypercharge as electric charge.
 
 If we now gauge hypercharge, by including $B_{\mu}$ in the covariant derivative and $g'\neq 0$, the diagonal subgroup $SU(2)_f$ breaks down to its $U(1)$ subgroup and becomes local. In this way a theory which is locally $U(1)_{\text{EM}}$ symmetric is obtained. 
 
 Since $U(1)_{\text{EM}}$ is a local symmetry, we again have to discuss gauge-invariance. However, as this is an Abelian gauge group, this operates differently. But due to the fact that every representation is one-dimensional, this can be solved like in QED \cite{Haag:1992hx,Lavelle:1995ty}, and is therefore not different than in the standard model \cite{Lavelle:1995ty}, and especially transparent to the FMS mechanism in the weak sector \cite{Maas:2017wzi}. It will therefore not be detailed here. Note that the gaugino does not carry charge and is thus $U(1)_{\text{EM}}$ gauge-invariant.

 Due to the introduction of $B_{\mu}$ and the breaking of the $W_{\mu}^a$ triplet we now have two fields in the neutral vector singlet channel which mix to create the $Z$ boson and the photon. This holds both at the elementary level and at the composite level. For the superpartners, the situation is similar: Introducing the bino $\ino{B}$ does not affect the charginos and merely mixes with the neutralinos. This implies that there are now two poles in the corresponding channels. Matching the new mass eigenstates with composite operators again reduces to a task of finding a suitable linear combination. 

 Still, this implies that the composite operators and the elementary ones do not differ in the hypercharge quantum numbers, but only in their gauge charges with respect to the unmixed part of the weak interactions.

\subsection{Multiple Generations and Quarks}
    \label{subsubsec:QCD}
    
    Including all three lepton generations substantially increases the complexity but changes nothing about our construction. In particular, generation indices are carried in the same way by elementary fields and composite fields. Intergeneration mixing is completely transparent to our composite operator construction as one could just introduce operators for each generation (c.f.~Tab.~\ref{tab:leptonToyModel_gaugeInvariantOperators})
\begin{align*}
    \dagg H L_e,~\dagg H L_{\mu},~\dagg H L_{\tau} && \rlepton_e,~\rlepton_{\mu},~\rlepton_{\tau}\\
    \dagg H \slepton_e,~\dagg H \slepton_{\mu},~\dagg H \slepton_{\tau} && \rslepton_e,~\rslepton_{\mu},~\rslepton_{\tau}.
\end{align*}
Both components of these operators are inherently gauge-invariant and can be linearly combined and rotated in generation space to augment all resulting mass eigenstates.

In the context of (S)QCD, the low energy description is all about objects which are built from elementary quarks and gluons to form colour neutral bound states, i.\ e.\ gauge-invariant with respect to $SU(3)_c$. Nevertheless, fields like the pion still carry $SU(2)_L$ charge which has to be taken care of. Luckily, in terms of electroweak and Higgs physics, the description of quarks is completely analogous to leptons \cite{Egger:2017tkd,Maas:2017wzi}, i.e. we can readily write down
\begin{align*}
    \Psi^{d} &= \begin{pmatrix}
      (\dagg HQ)_1 \\ v_u\bar{d}^{\dagger}
    \end{pmatrix} \simFMS v_u\begin{pmatrix}
        d \\ \bar{d}^{\dagger}
      \end{pmatrix} = v_u\psi^d\\
    \Psi^{u} &=\begin{pmatrix}
      (\dagg HQ)_2 \\ v_d\bar{u}^{\dagger}
    \end{pmatrix} \simFMS v_d\psi^{u},
\end{align*}
in analogy to Eq.~\eqref{eq:electron_neutrino_operators_unequal_vev} for leptons. The only difference is, that $\Psi^{u,d}$ are not yet physical as they still carry colour charge. Even though these \q{quark-Higgs bound states} cannot exist in isolation, such contractions are still important when building the usual colour singlets. An inherently gauge-invariant operator for $\pi^+$ would e.g. be \cite{Maas:2017wzi}
\begin{equation}
    \label{eq:pion}
    \begin{aligned}
        \Pi^+ \equiv \bar{\Psi}^d\Psi^u = \begin{pmatrix}
            v_u\bar{d} & (\dagg HQ)^{\dagger}_1
        \end{pmatrix}\begin{pmatrix}
            (\dagg H Q)_2 \\ v_d\bar{u}^{\dagger}
        \end{pmatrix} \\
        \simFMS v_uv_d\left(\bar{d}u +d^{\dagger}\bar{u}^{\dagger}\right) \sim v_uv_d\pi^+.
    \end{aligned} 
\end{equation}
Notice that the expression on the right hand side is a colour singlet but not an $SU(2)_L$ singlet which makes it apparent that using composite operators is also important in the (S)QCD subsector. Finally, just like before, we can use the $U(1)$ subgroup of $\custsym$ as well as the hypercharge assignments of the quarks in the SM to find that $\Psi^{d,u}$  carry electric charges $-1/3$ and $2/3$, respectively. Consequently, $\Pi^+$ carries the correct electric charge of $+1$.

Squarks and gluinos are not colour-gauge-invariant fields either and have to be treated in a bound state language with respect to $SU(3)_c$, too. Nevertheless, this has no effect on our construction. Gluinos carry no weak charge and are therefore only subject to confinement by the strong interaction. \q{Left-handed} squarks get an appropriate Higgs dressing, just like in the leptonic case.

\section{Summary}
  \label{sec:summary}

  We have shown that the MSSM, as the SM \cite{Maas:2017wzi,Frohlich:1980gj,Frohlich:1981yi,Egger:2017tkd} and the 2HDM \cite{Maas:2016qpu}, does not experiences any changes in its spectrum once physical composite states are used, rather than the elementary ones. As in the SM \cite{Maas:2020kda}, it is unlikely that this will change beyond our tree-level calculations. As we performed the analysis keeping the explicit SUSY breaking terms in a suitable way, it follows immediately that SUSY, and $R$-parity, is transparent to APT. However, this is neither manifest nor trivial, see appendix \ref{a:susy}. 
  
  While we cannot offer a proof at the current time, the observed structure allows us to conjecture that for a non-Abelian theory the question, whether the physical spectrum and tree-level spectrum are mapped onto each other in a one-to-one fashion has the same answer in the original theory, and its supersymmetrized version. This should also hold when SUSY breaks, as long as the SUSY breaking is not tampering with the gauge symmetry, as is the case, e.\ g., in gauge-mediated SUSY-breaking scenarios. These cases will require further scrutiny. In the case of gauged SUSY, the situation can be expected to be very different, and in fact SUSY may be completely absent from the physical spectrum \cite{Maas:2023emb}.

  What goes into this conjecture is that the weakly-charged elementary spectrum is not changed due to the presence of SUSY. Beyond the weak sector, the strong sector and its possibility for new hybridization and other non-perturbative alterations of the spectrum of supersymmetric multiplets, needs to be checked as well. Though this appears unlikely having impact on the experimentally accessible low-energy spectrum of the MSSM, this is a question of principle.

  Due to the mapping of the mass spectrum, this implies that the usual resolution of the hierarchy problem by SUSY \cite{Aitchison:2007fn} is not affected by using composite operators. In fact, as was already observed for the SM in \cite{Maas:2017wzi,Maas:2020kda}, the whole approach is at worst transparent to the hierarchy problem and fine-tuning issues. In fact, in a very precise sense \cite{Maas:2020kda}, it can actually reshape the problems as it maps it on gauge-invariant quantities.

  In total, the MSSM fits into the pattern so far observed that manifest gauge invariance does not lead to qualitative changes in the spectrum if the global group carried by the Higgs is at least as large as the gauge group. This is good news to phenomenology, as this implies that, up to the additional (sub-leading) terms in APT, predictions for experiment in the MSSM remain valid.
  
  Conversely, this implies that measurements will be affected in the same way as for the SM. E.\ g., high-energetic production of sfermions pairs will have different electroweak corrections in APT than in PT, due to cancellations of virtual and real electroweak radiation \cite{Maas:2022gdb}. Likewise, in some MSSM processes additional contributions may be suppressed at the NNLO level, as the chiral structure is the same as in the SM \cite{Maas:2024hcn}. But there should be MSSM processes in which the suppression is at most NLO compared to the leading contribution \cite{Maas:2024hcn,Maas:2022gdb,Jenny:2022atm}. There is, however, an entirely different question to be answered: Can the contribution described here mask or mock-up the effects of the MSSM if there is only the SM? Investigations, e.\ g.\ into anomalous couplings, could be interpreted that such an effect is possible at least for some energy range \cite{Maas:2018ska}.\\

  \noindent{\bf Acknowledgements} \\

  This work was done within the scope of the FCC Feasibility Study.

  \appendix

  \section{Manifest SUSY}\label{a:susy}

  Attempting to use APT in a manifest SUSY-invariant way leads to the following problem. The introduction of a Higgs vev is a gauge choice, but necessarily one which makes supersymmetry not manifest. This follows as the Higgs field is part of a superspin multiplet. Introducing a condensate only in one component necessarily hides the supermultiplet structure. Consider the Higgs superspin multiplet $\Sigma=(\psi,\phi,\overline{\psi})$, which is build from two chiral supermultiplets to give a fundamental representation of the weak $SU(2)$. As has been seen in the main text, this structure of two chiral supermultiplets carries the $\custsym$ symmetry.

  A manifestly gauge-invariant and SUSY-invariant composite scalar\footnote{Note that any manifest SUSY-invariant operator is necessarily a superspin scalar.} operator would be $\Sigma^\dagger(x)\Sigma(x)$. Fixing a gauge with BEH effect and corresponding splitting of the Higgs field would yield for the operator
  \begin{equation}
      \Sigma^\dagger(x)\Sigma(x)=v\left(n^\dagger\eta(x)+\eta^\dagger(x)n\right)+\eta^\dagger(x)\eta(x)+\psi^\dagger\psi+\overline{\psi}^\dagger\overline{\psi}\nonumber
  \end{equation}
  Thus, the conventional APT reduction of section \ref{sec:apt} will only work for the Higgs-component, and it will look as if there was no other pole in this channel. However, applying SUSY before gauge-fixing to the same composite operator necessarily implies that all components have the same mass. Thus, the two two-fermion operators need to carry the same pole, and thus are meson-like bound states, yielding the correct number of degenerate states forming the SUSY multiplet.
  
  The situation would be similar for other supermultiplets. E.\ g., for a mass multiplet for leptons $\Omega$, a suitable operator would be $\Omega^\dagger\Sigma+\text{cc}$, where again only one component will have a mass pole from tree-level APT, and the other components will have a mass pole by virtue of SUSY as non-trivial bound states.
  
  While this is not in contradiction to APT, it invalidates the simplicity of just using tree-level APT. On the other hand, if SUSY is explicitly broken, like in the MSSM, this is no longer necessary, and the situation simplifies as in the main part of the paper. Alternatively, a gauge-dependent diagonal subgroup of the gauge symmetry and the supersymmetry could be used, instead, like using the diagonal subgroup of $\custsym$ and gauge symmetry in the main text. Also, a formulation without manifest SUSY likewise works. Conversely, if one would like to keep global supersymmetry manifest at the gauge- fixed level, this forbids a gauge choice implementing a BEH effect. This is not surprising, given the links between both symmetries already in PT \cite{Aitchison:2007fn,Weinberg:2000cr}.
   
   A common treatment of SUSY and the BEH effect will only be possible if both symmetries are of the same nature. If both are global, this is the usual situation well known in standard approaches, and APT will not be applicable nor necessary. On the other hand, making SUSY a gauge symmetry, both are again on the same footing. This is then supergravity with an additional local gauge symmetry \cite{Freedman:2012zz}. In that case, just as in ordinary quantum gravity \cite{Maas:2019eux}, APT appears to be possible \cite{Maas:2023emb}. But in that case there will be two BEH effects, one on the level of the Higgs field, and one on the level of the metric. Then both symmetries are no longer manifest, and again APT is directly applicable for both, like this is already the case in ordinary quantum gravity \cite{Maas:2019eux}.

\bibliographystyle{bibstyle}
\bibliography{bib}

\begin{thebibliography}{10}

\bibitem{Bohm:2001yx}
M.~B\"ohm, A.~Denner, and H.~Joos,
\newblock {\em {Gauge theories of the strong and electroweak interaction}}
  (Teubner, Stuttgart, 2001).

\bibitem{pdg}
Particle Data Group, R.~L. Workman {\em et~al.},
\newblock PTEP {\bf 2022}, 083C01 (2022).

\bibitem{Elitzur:1975im}
S.~Elitzur,
\newblock Phys. Rev. {\bf D12}, 3978 (1975).

\bibitem{Lee:1974zg}
B.~Lee and J.~Zinn-Justin,
\newblock Phys.Rev. {\bf D5}, 3137 (1972).

\bibitem{Frohlich:1980gj}
J.~Fr\"ohlich, G.~Morchio, and F.~Strocchi,
\newblock Phys.Lett. {\bf B97}, 249 (1980).

\bibitem{Frohlich:1981yi}
J.~Fr\"ohlich, G.~Morchio, and F.~Strocchi,
\newblock Nucl.Phys. {\bf B190}, 553 (1981).

\bibitem{Englert:2004yk}
F.~Englert,
\newblock Proceedings: 50 years of Yang-Mills theory , 65 (2005),
  hep-th/0406162.

\bibitem{Englert:2014zpa}
F.~Englert,
\newblock Rev. Mod. Phys. {\bf 86}, 843 (2014).

\bibitem{Maas:2012ct}
A.~Maas,
\newblock Mod. Phys. Lett. {\bf A27}, 1250222 (2012), 1205.0890.

\bibitem{Fujikawa:1978fu}
K.~Fujikawa,
\newblock Prog. Theor. Phys. {\bf 61}, 627 (1979).

\bibitem{Gribov:1977wm}
V.~N. Gribov,
\newblock Nucl. Phys. {\bf B139}, 1 (1978).

\bibitem{Singer:1978dk}
I.~M. Singer,
\newblock Commun. Math. Phys. {\bf 60}, 7 (1978).

\bibitem{Banks:1979fi}
T.~Banks and E.~Rabinovici,
\newblock Nucl.Phys. {\bf B160}, 349 (1979).

\bibitem{Shifman:2012zz}
M.~Shifman,
\newblock {\em {Advanced topics in quantum field theory: A lecture course}}
  (Cambridge University Press, 2012).

\bibitem{Maas:2017wzi}
A.~Maas,
\newblock Progress in Particle and Nuclear Physics {\bf 106}, 132 (2019),
  1712.04721.

\bibitem{Maas:2020kda}
A.~Maas and R.~Sondenheimer,
\newblock Phys. Rev. D {\bf 102}, 113001 (2020), 2009.06671.

\bibitem{Dudal:2020uwb}
D.~Dudal {\em et~al.},
\newblock Eur. Phys. J. C {\bf 81}, 222 (2020), 2008.07813.

\bibitem{Maas:2023emb}
A.~Maas,
\newblock {\em Rigorous Trails Across Quantum and Classical Physics: A Volume
  in Tribute to Giovanni Morchio} (Springer, 2023), chap. {The
  Fr\"ohlich-Morchio-Strocchi mechanism: A underestimated legacy}, 2305.01960.

\bibitem{Jenny:2022atm}
P.~Jenny, A.~Maas, and B.~Riederer,
\newblock Phys. Rev. D {\bf 105}, 114513 (2022), 2204.02756.

\bibitem{Maas:2018ska}
A.~Maas, S.~Raubitzek, and P.~T\"orek,
\newblock Phys. Rev. {\bf D99}, 074509 (2019), 1811.03395.

\bibitem{Maas:2023nsa}
A.~Maas,
\newblock {Experimental signatures of subtleties in the Brout-Englert-Higgs
  mechanism},
\newblock in {\em {57th Rencontres de Moriond on Electroweak Interactions and
  Unified Theories}}, 2023, 2305.07395.

\bibitem{Maas:2022gdb}
A.~Maas and F.~Reiner,
\newblock Phys. Rev. D {\bf 108}, 013001 (2023), 2212.08470.

\bibitem{Maas:2016ngo}
A.~Maas and P.~T\"orek,
\newblock Phys. Rev. {\bf D95}, 014501 (2017), 1607.05860.

\bibitem{Maas:2017xzh}
A.~Maas, R.~Sondenheimer, and P.~T\"orek,
\newblock Annals of Physics {\bf 402}, 18 (2019), 1709.07477.

\bibitem{Maas:2018xxu}
A.~Maas and P.~T\"orek,
\newblock Annals Phys. {\bf 397}, 303 (2018), 1804.04453.

\bibitem{Dobson:2022crf}
E.~Dobson, A.~Maas, and B.~Riederer,
\newblock PoS {\bf LATTICE2022}, 210 (2022), 2211.16937.

\bibitem{Sondenheimer:2019idq}
R.~Sondenheimer,
\newblock Phys. Rev. D {\bf 101}, 056006 (2020), 1912.08680.

\bibitem{Maas:2016qpu}
A.~Maas and L.~Pedro,
\newblock Phys. Rev. {\bf D93}, 056005 (2016), 1601.02006.

\bibitem{Aitchison:2007fn}
I.~Aitchison,
\newblock {\em {Supersymmetry in Particle Physics. An Elementary Introduction}}
  (Cambridge University Press, 2007).

\bibitem{Weinberg:2000cr}
S.~Weinberg,
\newblock {\em {The quantum theory of fields. Vol. 3: Supersymmetry}}
  (Cambridge, UK: Univ. Pr., 2000).

\bibitem{Kalka:1997us}
H.~Kalka and G.~Soff,
\newblock {\em {Supersymmetrie}} (Teubner (Dresden), 1997).

\bibitem{Martin:1997ns}
S.~P. Martin,
\newblock Adv. Ser. Direct. High Energy Phys. {\bf 18}, 1 (1998),
  hep-ph/9709356.

\bibitem{Schreiner:2022ms}
P.~Schreiner,
\newblock The manifestly gauge-invariant spectrum of the minimal supersymmetric
  standard model,
\newblock Master's thesis, University of Graz, Austria, 2022.

\bibitem{Egger:2017tkd}
L.~Egger, A.~Maas, and R.~Sondenheimer,
\newblock Mod. Phys. Lett. {\bf A32}, 1750212 (2017), 1701.02881.

\bibitem{Maas:2012tj}
A.~Maas,
\newblock Mod.Phys.Lett. {\bf A28}, 1350103 (2013), 1205.6625.

\bibitem{MPS:unpublished}
A.~Maas, S.~Pl\"atzer, and R.~Sondenheimer,
\newblock unpublished.

\bibitem{Wurtz:2013ova}
M.~Wurtz and R.~Lewis,
\newblock Phys.Rev. {\bf D88}, 054510 (2013), 1307.1492.

\bibitem{Dobson:2025kcx}
E.~Dobson, A.~Maas, S.~Pl\"atzer, and B.~Riederer,
\newblock (2025), 2501.19212.

\bibitem{Maas:2015gma}
A.~Maas,
\newblock Mod. Phys. Lett. {\bf A30}, 1550135 (2015), 1502.02421.

\bibitem{Caudy:2007sf}
W.~Caudy and J.~Greensite,
\newblock Phys. Rev. {\bf D78}, 025018 (2008), 0712.0999.

\bibitem{Dobson:2022ngz}
E.~Dobson, A.~Maas, and B.~Riederer,
\newblock Annals Phys. {\bf 457}, 169404 (2023), 2211.05812.

\bibitem{Maas:2013aia}
A.~Maas and T.~Mufti,
\newblock JHEP {\bf 1404}, 006 (2014), 1312.4873.

\bibitem{Philipsen:1996af}
O.~Philipsen, M.~Teper, and H.~Wittig,
\newblock Nucl.Phys. {\bf B469}, 445 (1996), hep-lat/9602006.

\bibitem{Dolan:1974gu}
L.~Dolan and R.~Jackiw,
\newblock Phys. Rev. {\bf D9}, 2904 (1974).

\bibitem{Nielsen:1975fs}
N.~K. Nielsen,
\newblock Nucl. Phys. {\bf B101}, 173 (1975).

\bibitem{Maas:2024hcn}
A.~Maas, D.~M. van Egmond, and S.~Pl\"atzer,
\newblock PoS {\bf ICHEP2024}, 056 (2025), 2409.20131.

\bibitem{Afferrante:2020hqe}
V.~Afferrante, A.~Maas, and P.~T\"orek,
\newblock Phys. Rev. D {\bf 101}, 114506 (2020), 2002.08221.

\bibitem{DeGrand:2006zz}
T.~DeGrand and C.~E. Detar,
\newblock {\em {Lattice methods for quantum chromodynamics}} (World Scientific,
  New Jersey, 2006).

\bibitem{Itzykson:1980rh}
C.~Itzykson and J.~Zuber,
\newblock {\em Quantum field theory}International Series In Pure and Applied
  Physics (McGraw-Hill, 1980).

\bibitem{Piguet:1995er}
O.~Piguet and S.~P. Sorella,
\newblock {\em {Algebraic renormalization: Perturbative renormalization,
  symmetries and anomalies}}volume~28 (, 1995).

\bibitem{Dudal:2021dec}
D.~Dudal, D.~M. van Egmond, I.~F. Justo, G.~Peruzzo, and S.~P. Sorella,
\newblock Phys. Rev. D {\bf 105}, 065018 (2022), 2111.11958.

\bibitem{Gunion:1984yn}
J.~F. Gunion and H.~E. Haber,
\newblock Nucl. Phys. B {\bf 272}, 1 (1986).

\bibitem{Grzadkowski:2010dj}
B.~Grzadkowski, M.~Maniatis, and J.~Wudka,
\newblock JHEP {\bf 11}, 030 (2011), 1011.5228.

\bibitem{Branco:2011iw}
G.~Branco {\em et~al.},
\newblock Phys.Rept. {\bf 516}, 1 (2012), 1106.0034.

\bibitem{Haag:1992hx}
R.~Haag,
\newblock {\em {Local quantum physics: Fields, particles, algebras}} (Springer,
  Berlin, 1992).

\bibitem{Lavelle:1995ty}
M.~Lavelle and D.~McMullan,
\newblock Phys. Rept. {\bf 279}, 1 (1997), hep-ph/9509344.

\bibitem{Freedman:2012zz}
D.~Z. Freedman and A.~Van~Proeyen,
\newblock {\em {Supergravity}} (Cambridge Univ. Press, Cambridge, UK, 2012).

\bibitem{Maas:2019eux}
A.~Maas,
\newblock SciPost Phys. {\bf 8}, 051 (2020), 1908.02140.

\end{thebibliography}

\end{document}